\documentclass[twocolumn]{revtex4}

\usepackage{graphicx}
\usepackage{latexsym}

\begin{document}


\title{Finite-Width Bundle is Most Stable in a Solution with Salt} 



\author{Takuya Saito}
\thanks{Electronic mail: saito@stat.phys.kyushu-u.ac.jp}
\affiliation{Department of Physics, Kyushu University, Fukuoka 812-8581, Japan}

\author{Kenichi Yoshikawa}
\affiliation{Department of Physics, Graduate School of Science, Kyoto University, Kyoto, 606-8502, Japan and \\ Spatio-Temporal Order Project, ICORP, JST, Kyoto University, Kyoto 606-8502, Japan}


\date{\today}

\begin{abstract}
We applied the mean-field approach to a columnar bundle assembled by the parallel arrangement of stiff polyelectrolyte rods in a salt bath.  The electrostatic potential can be divided into two regions: inside the bundle for condensed counter-ions, and outside the bundle for free small ions.  To determine the distribution of condensed counter-ions inside the bundle, we use a local self-consistent condition that depends on the charge density, the electrostatic potential, and the net polarization. The results showed that, upon bundle formation, the electric charge of polyelectrolytes,  even those inside the bundle, tend  to survive in an inhomogeneous manner, and thus their width remains finite under thermal equilibrium because of the long-range effect of charge instability.
\end{abstract}


\maketitle 



%
%

%

\def\degC{\kern-.2em\r{}\kern-.3em C}

\section{Introduction}

For electrically charged soft materials including biological polyelectrolytes such as DNA and actin, it is important to understand their behavior as a many-body Coulomb system.  A well-known model under a mean field approximation is the Poisson-Boltzmann equation (PB eq.), which has often been used to describe the distribution of small ions around a stiff polyelectrolyte rod in a good solvent~\cite{MacGillivray_01_1972}. A typical example is negatively charged DNA with a persistence length of $\sim$ 50~nm.  The PB eq. allows one to estimate an ion distribution not only for a short rigid fragment of DNA, but also for a segment of a long coiled DNA shorter than the persistence length.  However, we have to be very careful when interpreting the results of the PB for polyelectrolytes in a poor solvent.

 DNA condensation is induced by various condensing agents~\cite{PRONAS_Lerman_1971,Wilson_Bloomfield_1979,Widom_1980} such as multivalent cations, neutral polymers (polyethylene glycol, etc), and cationic surfactants.  The experimental observation of the aggregation of short oligomeric DNA suggests that their local structure is an aligned ordered packing~\cite{Livolant_2005}.  Furthermore, this applies not only to condensed DNA, which consists of the multiple-assembly of short fragments, but also to compact DNA, which consists  of a single long chain.  Upon the addition of condensing agents, long DNA discretely folds from a coil to a compact state at the level of a single molecule. This DNA compaction can be characterized as a first-order phase transition under Landau's argument~\cite{YoshikawaPRL_1996}. The typical morphology of compact DNA is a toroid, the local structure of which can be regarded as an ordered bundle based on the parallel arrangement of its segments, when we note the cross-section of a toroid cut with the plane including the rotational symmetric axis.

The transition to a parallel packed arrangement should lead to a drastic change in the distribution of small ions. The condensation of counter ions inside the packed state, which accompanies both the bundling of short DNAs and the folding of a long DNA chain, should be enhanced to diminish coulombic instability. In fact, the experimental results support this significant change~\cite{BiophysJ_Yamasaki_2001}.  The condensed counter-ions in compact DNA have a notable contribution, but this is not appropriately estimated by the PB eq.  This implies that a new approach that is different than the PB eq. is needed for this case.

In this article, we adopt a different approach from the PB eq. for a bundle that consists of  the parallel-arrangement packing of stiff polyelectrolyte rods in the presence of salt. The morphology also includes a column with a finite diameter, since the experimentally observed size is finite.  For example, the bundle assembled by stiff rod-like polyelectrolytes such as F-actins has a finite width~\cite{PRE_Hosek_Tang_2004}.  Also, in the case of a long single DNA chain, it has been found that the coil and compact states coexist in a single molecule under some conditions~\cite{EPL_Iwaki_2004}.  In addition, at high concentrations of long DNA, a stiff thick bundle with a persistence length on the order of $\mu m$ by multiple-assembly has been observed~\cite{JCP_Iwataki_2004}.  At present, it is difficult to experimentally determine whether these phenomena regarding a finite size arise from the free energy minimum in the equilibrium condition or a  kinetic effect. In this study, we obtain the minimum point at a finite diameter on the free energy profile, and discuss the mechanism.

\section{columnar bundle assembled by the parallel arrangement of stiff polyelectrolyte rods in a salt bath}

 Imagine a columnar bundle that is formed by the parallel arrangement of stiff polyelectrolyte rods,  as shown in Fig~1.  A stiff polyelectrolyte rod carries charge $-Q\,(<0)$ and has a line density of charge $-q/d$ ($q$ is the unit charge), which means that the length of the rods is $L \,(=Qd)$.
 A bundle composed of these rods is dissolved in the presence of a monovalent salt ($1+$,$1-$), the concentration of which is $c$.  The intrinsic dielectric constant in solvent (water) is $\epsilon_0$.  Also, we introduce the  Bjerrum length $l_{\rm{B}} \equiv q^2 / 4 \pi \epsilon_0 k_{\rm{B}} T \equiv q^2 \beta / 4 \pi \epsilon_0  $ ($k_{\rm{B}} $ is the Boltzmann constant, $T$ is the thermal temperature, and $\beta = 1/ k_{\rm{B}} T$). 

\begin{figure}[t!]
\begin{center}
\includegraphics[scale=0.45]{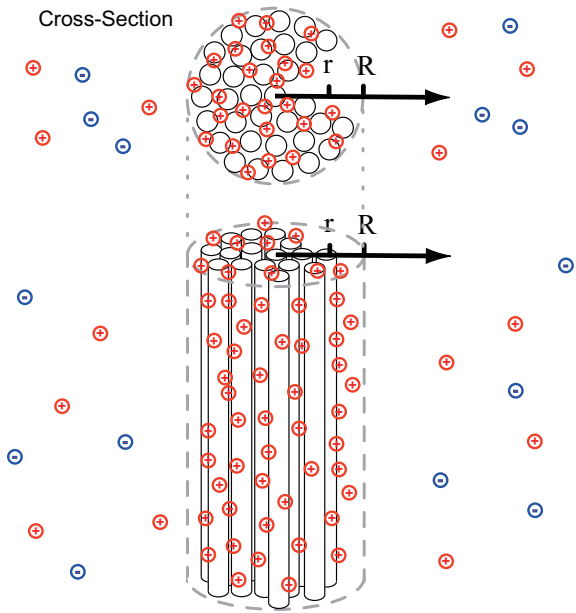}
\caption{(Color Online) Schematic representation of a columnar bundle assembled by the parallel arrangement of stiff polyelectrolyte rods. Monovalent counter-ions (monovalent cations) condense into the bundle, while counter-ions and co-ions are in the bulk. The remaining electric charge of the bundle is screened by the surrounding small ions in bulk.  $r$ is the distance from the central axis of the columnar bundle, and $R$ is the bundle radius.
}
\label{fig1}
\end{center}
\end{figure}

\begin{figure}[h!]
\begin{center}
\includegraphics[scale=0.70]{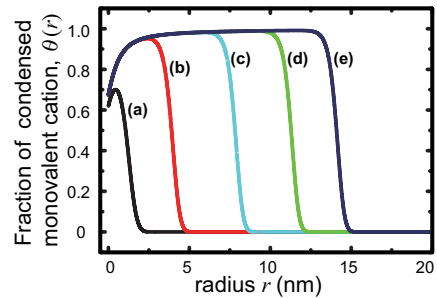}
\caption{(Color Online) Fraction of condensed counter-ions per polyelectrolyte $\theta (r)$ as a function of the radial distance from the central axis of the bundle $r$.  This profile is obtained from the differential equation (Eq.~(\ref{diff})) at various values of electro-neutrality at the central axis $\theta (r=0)$. (a)~$\theta_{\rm a}$, (b)~$\theta_{\rm b}$, (c)~$\theta_{\rm c}$, (d)~$\theta_{\rm d}$, (e)~$\theta_{\rm e}$. ($\theta_{\rm a}<\theta_{\rm b}<\theta_{\rm c}<\theta_{\rm d}<\theta_{\rm e}$).  See reference~\cite{theta_0} for these values.  The parameters are $d=0.17~\rm{nm}$, $l_{\rm{B}}=0.71~{\rm{nm}}$, $a=1~{\rm{nm}}$, $v=1~\rm{nm^3}$, $\lambda= 0.1~\rm{nm}$.}
\label{fig2}
\end{center}
\end{figure}

In this system, monovalent counter-ions can not induce the bundling of polyelectrolytes independently without any additional attraction. As an additional attractive force, we consider non-electrostatic forces such as the depletion force by a neutral polymer,  etc.  This situation corresponds to DNA compaction induced by polyethylene glycol (PEG).

To formulate this system, we use the concept of the Oosawa-Manning condensation theory (OM theory), which has been applied to a dispersed polyelectrolyte~\cite{OosawaBook, Manning_review} by dividing the potential into two regions:  inside and outside the polyelectrolyte.  This theory is superior to the PB eq. not only because it is easy to understand, but also because it corresponds to the experimental results.  Analogously, we divide the system into two regions, with small ions (counter-ions and co-ions) outside the bundle, and condensed counter-ions inside the bundle.

Here, we consider the self-consistent local condition for the interior ion distribution, by relating $D(r)$, $E(r)$ and $P(r)$ in the following relation:  
\begin{eqnarray}
D(r) &=& \epsilon_0  E(r) +  P(r),
\label{total}
\end{eqnarray}
where $D(r)$ is the electric displacement field, $E(r)$ is the electric field, $P(r)$ is the polarization, and $r$ is the radial coordinate perpendicular at the central axis of the  bundle, as shown in Figure~\ref{fig1}.  Hereafter,  we regard $D (r)$, $E (r)$ and $P (r)$ as the "{\it mean}" and "{\it continuous}" field on a coarse-grain scale.  The cross-sectional density of polyelectrolyte rods in the bundle are also uniform, and the cross-sectional area per polyelectrolyte is $v/d$.  The condensed small ions into the bundle are monovalent cations, the fraction of which is defined as $\theta (r)$ per polyelectrolyte rod.  In the following discussion, we consider that $0 \leq \theta (r) \leq 1$.

\subsubsection{Electric field: $E(r)$}
In the equilibrium state, the chemical potential is equivalent everywhere inside the bundle,  as follows:

\begin{eqnarray}
\mu_{\rm{ele}}^{\rm{in}}(r) + \mu_{\rm{tra}}^{\rm{in}}(r) = {\rm constant }.
\label{relation_interior_chem}
\end{eqnarray}
where $\mu_{\rm{ela}}^{\rm{in}}(r)$ is the chemical potential of the mean electric field inside the bundle and $\mu_{\rm{tra}}^{\rm{in}}(r)$ represents the contribution of translational entropy.  Here, we neglect the higher-order multipole composition.  The electrostatic chemical potential is written as $\mu_{\rm{ele}}^{\rm{in}}(r) = q \phi (r)$.  The chemical potential for the translational entropy of a  monovalent counter-ion is also given by $\mu_{\rm{tra}}^{\rm{in}}(r) = k_{\rm{B}} T \log{\left( \theta(r) /c v \right)}$.  Therefore, the derivative of the total chemical potential with respect to $r$ is $\partial (\mu_{\rm{ele}}^{\rm{in}}(r)) + \mu_{\rm{tra}}^{\rm{in}}(r) )/ \partial r=0$.  Thus, this leads to

\begin{eqnarray}
q \beta E(r) =  \partial_{r} (\log{\frac{\theta(r)}{ c v}}) = \partial_{r} (\log{\theta (r)}),
\label{electric}
\end{eqnarray}
where the mean electric field $E(r) \equiv - \partial \phi (r) / \partial r$.

\subsubsection{Polarization: $P (r)$ }
The degree of polarization should be closely related to the excluded volume of polyelectrolytes and counter-ions.  Counter-ions should condense on the surface of each polyelectrolyte rod in the bundle. We may assume that the separation of the two charges of the dipole is the closest-approach distance.  In this case, the polarization density is considered to be

\begin{eqnarray}
P(r) =   - q \lambda \theta (r) / v,~~(\lambda >0),
\label{polar}
\end{eqnarray}
where $\lambda$ is the effective mean approach between the polyelectrolyte and counter-ions.

\subsubsection{Electric displacement field:  $D(r)$}
From the definition of $\theta (r)$, the charge density is given by
\begin{eqnarray}
{\rm div} D(r) = -q (1 - \theta (r)) / v.
\label{displace}
\end{eqnarray}

\subsubsection{Local condition inside the bundle}
${\rm div} D (r) = \epsilon_0 {\rm div} E (r) + {\rm div} P(r) $ is given by the divergence in Eq~(\ref{total}). If we include Eqs.~(\ref{electric}), (\ref{polar}) and (\ref{displace}), we obtain the local condition of the ion distribution in the bundle as follows:

\begin{eqnarray}
- l_{\rm{B}} r (1 - \theta (r)) &=& \frac{v}{4 \pi} \partial_{r} \{ r \partial_{r} ( \log{ \theta(r) } ) \} +  l_{\rm{B}}  \partial_{r} (- r \lambda \theta (r)).\nonumber\\
\label{diff}
\end{eqnarray}

\subsubsection{Boundary condition at the central axis ($r=0$)}
In the above equation, there are two conditions at the central axis ($r=0$) and  the bundle surface ($r=R$).  The boundary condition at the central axis can be obtained under the assumption that the displacement field inside is continuous everywhere.   In this case, the displacement field is zero at the central axis.  Therefore, $ D(r=0) =0$ leads to

\begin{eqnarray}
D(r =0)
&=& \epsilon_{0} E(0) + P(0)\nonumber\\
&=& \epsilon_{0} \frac{k_{\rm{B}} T}{q} \frac{ \partial_{r} \theta(r)|_{r=0}  }{ \theta(0) } - \frac{ q \lambda \theta (0)}{v}  =0.
\label{cent_condition}
\end{eqnarray}
where $ D_r (r)$ is the radial displacement field~\cite{DisField}.  Note that $D (r)$ is represented by $E(r)$ and $P(r)$ (\,Eqs.~(\ref{electric}), (\ref{polar})\,).

\begin{figure}[t!]
\begin{center}
\includegraphics[scale=0.60]{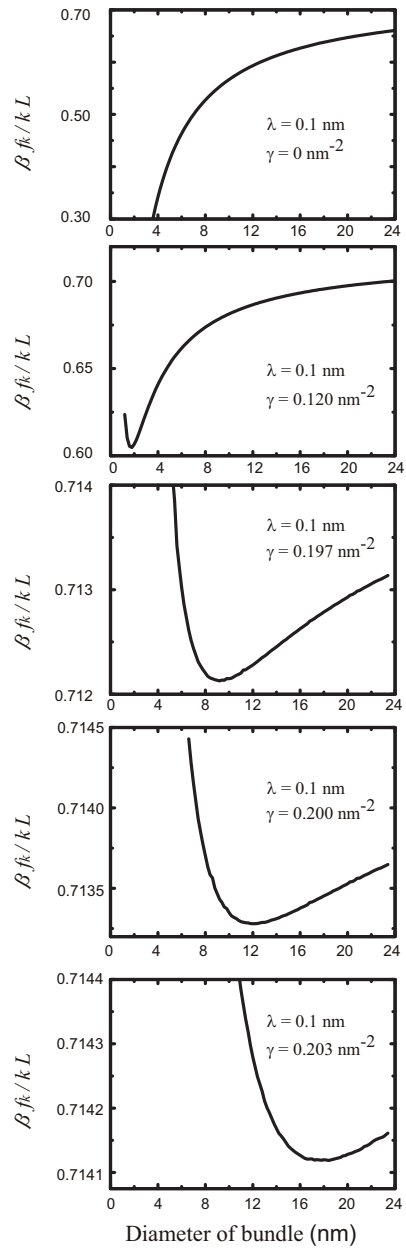}
\caption{The plot of the free energy $\beta f_{k} / k L $ as a function of the bundle diameter $2 R\,(=2 \sqrt{kv/\pi d})$ with changes in the value of the surface tension $\gamma$ at $\lambda = 0.1~{\rm nm}$.  $f_{k}/k$ corresponds to the free energy per stiff polyelectrolyte rod in the bundle with width $R$.  The parameters used in the calculation are $\gamma = 0$, $0.120$, $0.197$, $0.200$, and $0.203~{\rm nm}^{-2}$ at a salt concentration $c=0.3~{\rm M}$. The other parameters are the same as those in Fig.~\ref{fig2}.} 
\label{fig3}
\end{center}
\end{figure}

In addition, the boundary at $r=R $ will be naturally associated with the relation to the total surviving charge of the bundle, which is screened by the linearized PB-type ion atmosphere.  Note that the total surviving charge in the bundle will be determined by the free energy minimum as discussed later.

Figure~\ref{fig2} shows examples of solutions at various proportions of condensed counter-ion at the central axis, $\theta (r=0)$. The proportion of condensed counter-ion inside the bundle greatly decreases far from the central axis.

\subsection{Free energy of the bundle with a finite width}
In this section, we consider the bundle with a finite width, using the above local condition. First, let us consider the coexistence of gas and liquid in a finite-volume system that contains a finite number of particle, to emphasize the essential points regarding the finite-width phenomenon of a bundle. Gas and liquid coexist in the presence of some attraction such as van der Waals force.  In this situation, the finite cluster sizes arise from the penalty of translational entropy when the number of gas molecules in the system decreases.  However, the mechanism in the observed bundle with a finite width is clearly different.

 An entropic penalty is incurred along with bundle formation by the condensation of small ions into the bundle, and by the decrease in the degrees of freedom on polyelectrolyte rods.  However, a finite distribution of the bundle width is observed under the experimental conditions, where sufficient small ions are dissolved to act as a particle bath~\cite{PRE_Hosek_Tang_2004}, and where there is a high concentration of rods~\cite{Matsuzawa_2004}.  Therefore, to essentially understand the finite-width distribution of the bundle as an electrostatic effect, we must consider the mechanism, which should not be attributed to the effect of the finite number of polyelectrolyte rods or small ions.

In this system, the total free energy $F_{\rm{B-tot}}$ is written as
\begin{eqnarray}
F_{\rm{B-tot}} = F_{\rm{B-ent}} + F_{\rm{B-int}}.
\end{eqnarray}
$F_{\rm{B-ent}}$ corresponds to the degrees of freedom of the polyelectrolyte ( translational and rotational entropy for stiff rods, or the elastic free energy in the case of a semi-flexible chain).  $F_{\rm{B-int}}$ is the free energy of the effective interaction part between rods.  As mentioned above, we only examine the contribution that is essential for the finite-width effect $F_{\rm{B-int}}$.  This can be divided into different contributions, as follows:
\begin{eqnarray}
& &F_{\rm{B-int}} = \sum_{k} n_{k} f_{k} \nonumber\\
& &~~~~~~~~= \sum_{k} n_{k} \left( f_{k, \rm{ele-in}} + f_{k, \rm{scr}} + f_{k, \rm{tra}} + f_{k, \rm{atr}} \right), \\
& &\beta  f_{k, \rm{ele-in}} = L \int^{R}_{0} 2 \pi r {\rm d}r \frac{1}{2 k_{\rm B} T \epsilon_0} D_{r} (r) \left( D_{r} (r) -  P(r) \right),\nonumber\\
\label{mu_bind}
\nonumber\\
& &\beta  f_{k,\rm{scr}} = L \left( \int^{R}_{0} 2 \pi r {\rm d} r \frac{( 1 -   \theta (r))}{v} \right)^2 \frac{ l_B }{\kappa R }
\frac{K_0[ \kappa R]}{K_1[ \kappa R]},\nonumber\\
& &\beta  f_{k, \rm{tra}} = L \int^{R}_{0} 2 \pi r {\rm d} r \, ( \theta (r)/v ) \log{ (\theta(r) / e c v )},\nonumber\\
& &\beta  f_{k,\rm{atr}} = 2 \pi R L \gamma,
\label{our_model}
\end{eqnarray}
where $f_{k}$ is the free energy of the bundle formed by $k$ stiff polyelectrolyte rods, and $n_{k}$ is the number of bundles labeled by $k$ (total number of rods is $\sum_{k} k n_{k} $), and $k v = \pi R^2 d$.  $ f_{k,\rm{ele-in}} $ is the electrostatic free energy inside the polyelectrolyte bundle and $ f_{k,\rm{scr}}$ is the contribution of the ion atmosphere around the bundle.  The inverse Debye length is $\kappa = \sqrt{8 \pi l_{\rm B} c}$ and $K_{\nu}(x)$ are the modified Bessel functions of order $\nu$. $ f_{k,\rm{tra}}$ is the contribution from the translational entropy of the monovalent counter-ion. $ f_{k,\rm{atr}}$ is the additional attraction between polyelectrolyte rods, which is not attributable to small ions ($e.g.$, depletion force by neutral polymer). The ion atmosphere around the bundle is calculated as the linearized PB-type, like that around a single highly charged polyelectrolyte~\cite{CurOpi_Andelman_2004}.  Also, $\gamma$ is the surface tension (divided by $k_{\rm B} T$). We do not need to consider the volume part by attractive interaction, since that part of $ f_{k,{\rm atr}} /k $ is independent of $k$ under the assumption of pair-wise attraction. When we apply the ion distribution obtained from Eq.~(\ref{diff}), we minimize the free energy $ f_{k}/k $, which is the total free energy of one bundle $f_{k}$ divided by the constitutive number of polyelectrolyte rods.

\section{Results and Discussion}

 Figure~\ref{fig3} shows the profile for the free energy of one bundle divided by the constitutive number of polyelectrolyte rods and by its length $ \beta f_{k} /k L$, as a function of $R$, at various values of $\gamma$. The free energy $ f_{k}/k $ at each width $R$ (or $k$) is minimized by changing the degree of electro-neutrality at the central axis $\theta (r=0)$. Note that the free energy is divided by $k_{\rm B}T (=\beta^{-1})$ in this plot. As shown in Fig.~\ref{fig3}, when the surface tension $\gamma$ is greater, minimum points with a broad concave region appear. This indicates that bundles with a finite width are distributed.  Moreover, these shift to a larger radius $R$ with an increase in the surface tension $\gamma$.  

\begin{figure}[h!]
\begin{center}
\includegraphics[scale=0.65]{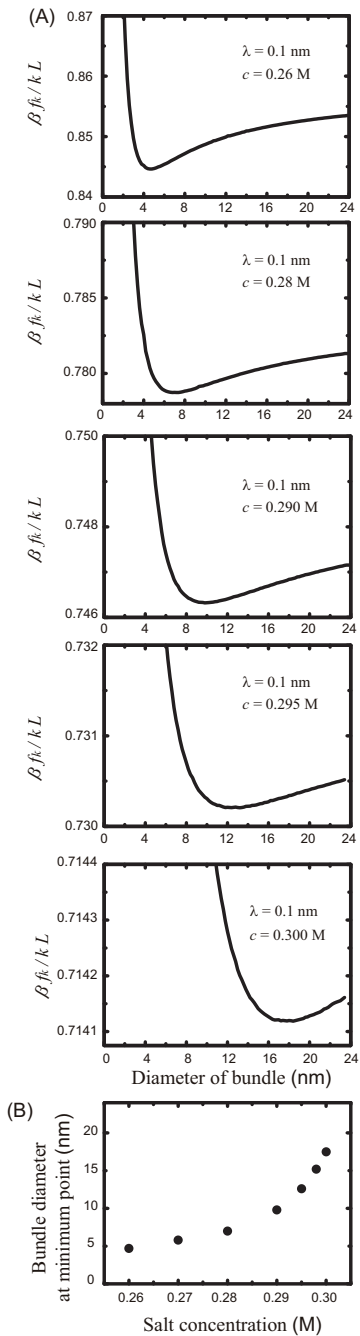}
\caption{ Dependence of the bundle diameter on the salt concentration. (A) The plot of free energy $\beta f_{k} / k L $ as a function of the bundle diameter $2R$ with changes in the salt concentration $c$ at a fixed surface tension $\gamma = 0.203~{\rm nm}^{-2}$.  The values for the salt concentration are $c = 0.260$, $0.280$, $0.290$, $0.295$, and $0.300~{\rm M}$. (B) Diagram of bundle diameters at minimum free energies as a function of the salt concentration. The other parameters used in the calculation are the same as those in Figs.~\ref{fig2}, \ref{fig3}.
}
\label{fig4}
\end{center}
\end{figure}

Also, Fig.~\ref{fig4}\,(A) shows the free energy plot $\beta f_{k} /k L$ as a function of the diameter $2R$ at various concentrations of salt $c$.  Figure~\ref{fig4}\,(B) shows the diagram of the bundle width at a minimum free energy $\beta f_{k} /k L$ as a function of the salt concentration $c$.  These results indicate that the minimum point shifts to a larger value when the salt concentration increases.

Note that, if the translational and rotational entropy of single rods, or the elastic free energy of a semi-flexible chain are added to the free energy $ f_{k} /k$, double minimum points should appear in  the free energy profile.  This means that finite-width bundles and single rods coexist in bulk, or folded and unfolded states coexist in a single chain.   In fact, in the case of long DNA, single-chain segregation has been experimentally observed in the presence of a neutral polymer such as PEG~\cite{EPL_Iwaki_2004}.

 Single-chain segregation has also been confirmed in the presence of multivalent cations~\cite{JCP_Iwaki_2008,Matsuzawa_2004}. Moreover, finite widths of a multiple aggregated assembly have been observed~\cite{Matsuzawa_2004}. These widths are close to those in our results, although they must be distinguished from PEG with regard to electrostatic effects.  This should not be a surprise, even with the different condensing agents, since one of the first factors is the excluded volume effects, which are comparable in both cases.   Furthermore, in the presence of a multivalent cation, the growth trend of width upon DNA compaction is consistent with those in their multiple assembly.  Similarly, this feature is expected to be valid for a PEG system, and our model satisfies this point.

How does the bundle have a finite width?  From the solution of the differential equation~(\ref{diff}), a charge in the interior of the bundle is not completely neutralized by monovalent counter-ions.  This remaining charge is almost determined by the competition between electrostatic free energy and translational entropy~\cite{EPL_Iwaki_2004}.  In fact, the free energy minimum of the bundle shifts to a greater width with an increase in the salt concentration as shown in Fig.~\ref{fig4}. This implies that the condensation of monovalent counter-ions into the bundle incurs a penalty in translational entropy, and this entropic penalty decreases at a higher salt concentration.

However, this contribution alone is not sufficient to discuss the finite-width effect.  The electrostatic penalty, including the translational entropy of small ions,  must increase slowly.  If the electrostatic penalty is immediately moderated with bundle growth before the decrease in the surface penalty, the bundle will continue to grow toward infinite width.  To clarify this point, we consider a simplified situation under the assumption that the bundle is uniformly neutralized by the ratio $\bar{\theta}$ without polarization.
 Electro-neutralization is almost completely determined by competition between the following two dominant contributions: 

\begin{eqnarray}
\beta  f_{k, {\rm ele}}^{P=0} &=& L ( \pi^2 l_{\rm B} /4   v^2 ) R^4 (1-\bar{\theta})^2
\\
\beta  f_{k, {\rm tra}}^{P=0} &=& L (\pi R^2 \bar{\theta} /v) \log{ ( \bar{\theta} / e c v ) },
\end{eqnarray}
where $f_{k, {\rm ele}}^{P=0}$ is the electrostatic free energy in the bundle, and $f_{k, {\rm tra}}^{P=0}$ is the contribution in the translational entropy of the counter-ion. Note that we use the electric field $E (r)= - r q (1-\bar{\theta}) /2 \epsilon_0 v$ obtained from Gauss's law in the calculation of $f_{k, {\rm ele}}^{P=0}$.  Here we expand it with respect to $\delta = 1- \bar{\theta}$, since a thick bundle is almost neutralized.   This minimization gives $\delta \simeq \log{(1/cv)}  / ( \pi l_{\rm B} R^2/ 2   v   + 1 ) \simeq 2v \log{(1/cv)} / (\pi l_{\rm B} R^{2}) $ at $R \gg \sqrt{ v/l_{\rm B} } $.   By substitution, we have $\beta  (f_{k, {\rm ele}}^{P=0} +  f_{k, {\rm tra}}^{P=0} ) /k L \simeq -  \log{(ecv)}/d - 3 ( \log{(cv)} )^2 v / \pi d l_{\rm B} R^2  $. The second term represents the free energy penalty with bundle growth.  On the other hand,  the penalty of the surface energy is $ \beta  f_{k,\rm{atr}}/k \sim \gamma /R $.  These sums $(f_{k, {\rm ele}}^{P=0} +  f_{k, {\rm tra}}^{P=0} + f_{k,\rm{atr}}) /k$ can have a {\it maximum} point, but not a {\it minimum} point in the free energy profile.  This indicates that the bundle grows toward infinite width.

Therefore, for the finite distribution of bundle width, a further effect of long-range instability should be required.  Our local condition in Eq.~(\ref{diff}) leads to non-uniform neutralization by the condensed counter-ions in the bundle,  which gives rise to slow-growth instability.  Moreover, the polarization $P(r)$ might reduce the electrostatic free energy loss in the integrand of Eq.~(\ref{mu_bind}) more than that loss in the case of invariant dielectric constant ($P(r)=0$).

Here, we emphasize the significance of a many-body effect on bundle formation.  As a model, we suppose a column morphology even in a thin bundle.  Under this assumption, the surface penalty is underestimated, or attraction is overestimated because the number of neighboring rods is overestimated.  This implies that, even if the two-body interaction is repulsive, many-body interaction could give an gain in free energy to form the bundle.  Although the finite-width bundle formed by stiff rods in the presence of multivalent cations has been argued to be due to a kinetic effect in the case of attractive two-body interaction between the parallel rods~\cite{Ha_1999}, at the same time bundle formation should be examined before the two-body interaction becomes attractive, since many-body interaction has advantageous effects by decreasing the surviving charge to reduce electrostatic repulsion, and by increasing the number of neighboring rods with bundle growth.

Before we conclude, we address the applicable scope of the present model.  The electric field is determined only by the translational entropy of condensed counter-ions,  as seen in Eq.~(\ref{electric}).  Therefore, this model can be used if the correlation by condensed counter-ions is not significant.  One possible example is chromatin fiber, if the quadrupolar interaction of the nucleosome structure is attractive and decoupled from condensed counter-ions. Although our proposed approach should include a certain level of contribution that is neglected in the simple PB~eq., this is similarly invalid, e.g., in the presence of a multivalent cation, where ion-exchange between monovalent and multivalent cations is expected to accompany the transition.  However, if the expression of the unconsidered local correlation is known and is added as a new term to the local condition in Eq.~(\ref{electric}), the coarse-grain electric field is given by the number of condensed counter-ions.  Thus, we hope that our approach can be expanded to various cases.

\section{Conclusion}

In conclusion, we propose the application of the mean field approach for a  bundle with a  parallel arrangement of stiff polyelectrolyte rods.  There is an inhomogeneously surviving charge of polyelectrolyte rods in the bundle interior, which is given by the local conditions which depend on the charge density, the electrostatic potential and the net polarization. The bundle width exhibits a finite distribution due to the long-range effect of electrostatic instability.  Furthermore, the stable finite width shifts to a larger value with an increase in the salt concentration.

~
\begin{acknowledgments}
We thank Dr. T. Sakaue at Kyushu University for a helpful discussion. This work was partly supported by Japan Society for the Promotion of Science (JSPS) under a Grant-in-Aid for Creative Scientific Research (Project No. 18GS0421), and by Ministry of Education, Culture, Sports, Science and Technology of Japan (No. 17076007).
\end{acknowledgments}

%



\begin{thebibliography}{10}
\bibitem{theta_0}%
  \BibitemOpen
  \bibinfo {note} {$\theta_{a}~=~0.62$, $\theta_{b}=0.67451$,\\
  $\theta_{c}=0.67451\,82659$,\\ $\theta_{d}= 0.67451\,82659\,16394$,\\
  $\theta_{e}= 0.67451\,82659\,16394\,12858$.\\ In this calculation, we utilize
  Runge-Kutta method by transforming it into simultaneous equations. In
  addition, Runge-Kutta-Fehlberg method is applied to the equation for the
  derivative of $\theta_r (r)$ depending on $\theta (r)$ and $\theta_r (r)$. In
  this case, tolerance limit is $10^{-26}$, and max step is $10^{-6}$. We start
  with $r=10^{-40}$ and step $10^{-40}$.}\BibitemShut{Stop}%
\bibitem{DisField}%
  \BibitemOpen
  \bibinfo {note} {In this calculation, we also apply the displacement field at
  arbitrary $r$ as follows: $D(r) = (q / 4 \pi l_{\rm B}) \partial_{r} ( \log{
  \theta(r) } ) - (q \lambda / v) \theta (r)$.}\BibitemShut{Stop}%
\end{thebibliography}

\end{document}